\documentclass[11pt,a4paper]{article}
\usepackage[acceptedWithA]{tacl2021v1}
\usepackage{environ}
\NewEnviron{acks}{\section*{Acknowledgments}\BODY}
\usepackage{amssymb}
\usepackage{enumitem}
\usepackage{amsmath}
\usepackage{algorithm}
\usepackage{textgreek}
\usepackage{algpseudocode}
\usepackage{tikz}
\usepackage{times}
\usepackage{latexsym}
\usepackage{booktabs}
\usepackage[T1]{fontenc}
\usepackage[utf8]{inputenc}
\usepackage{microtype}
\usepackage{inconsolata}
\usepackage[many,theorems]{tcolorbox}
\usepackage{float}

\newtcolorbox[auto counter]{examplebox}[2][]{
    breakable,
    colback=gray!5, colframe=gray!75!black,
    title=\textbf{Example \thetcbcounter: #2},
    #1
}

\usepackage{xspace,mfirstuc,tabulary}

\newif\iftaclinstructions
\taclinstructionsfalse 
\iftaclinstructions
\renewcommand{\confidential}{}
\renewcommand{\anonsubtext}{(No author info supplied here, for consistency with
TACL-submission anonymization requirements)}
\newcommand{\instr}
\fi

\iftaclpubformat 

\else

\fi

\title{\textsc{VeriRAG}: A Post-Retrieval Auditing of Scientific Study Summaries}

\author{
  Shubham Mohole$^1$, Hongjun Choi$^2$, Shusen Liu$^2$, Christine Klymko$^2$, \\
  \textbf{Shashank Kushwaha}$^3$, \textbf{Derek Shi}$^4$, \textbf{Wesam Sakla}$^2$, \textbf{Sainyam Galhotra}$^1$, \textbf{Ruben Glatt}$^2$
  \\
  $^1$Cornell University, USA \\
  $^2$Lawrence Livermore National Laboratory, USA \\
  $^3$University of Illinois Urbana‑Champaign, USA \\
  $^4$University of California, Los Angeles, USA
}

\date{}

\begin{document}
\maketitle
\begin{abstract}
Can democratized information gatekeepers and
 community note writers effectively decide what
 scientific information to amplify?  Lacking domain expertise, such
 gatekeepers rely on automated reasoning  agents that use RAG to ground evidence to cited sources.  But such standard RAG systems validate summaries via semantic grounding and suffer from ``methodological blindness,'' treating all cited evidence as equally valid regardless of rigor.  To address this, we introduce \textsc{VeriRAG}, a post-retrieval auditing framework that shifts the task from classification to methodological vulnerability detection. Using private Small Language Models (SLMs), \textsc{VeriRAG} audits source papers against the \textit{Veritable} taxonomy of statistical rigor. We contribute: (1)~a benchmark of 1,730 summaries with realistic, non-obvious perturbations modeled after retracted papers; (2)~the auditable \textit{Veritable} taxonomy; and (3)~an operational system that improves Macro F1 by at least 19 points over baselines using GPT-based SLMs, a result that replicates across MISTRAL and Gemma architectures. Given the complexity of detecting non-obvious flaws, we view \textsc{VeriRAG} as a ``vulnerability-detection copilot,'' providing structured audit trails for human editors. In our experiments, individual human testers found over 80\% of the generated audit trails useful for decision-making. We plan to release the dataset and code to support responsible science advocacy.
\end{abstract}
\section{Introduction}
Today, science journalists and social media moderators act as `amplification stations' \cite{kasperson1988social}, where poorly vetted studies erode public trust \cite{ophir2021effects}. While high-profile retractions highlight this danger, expert fact-checking remains unscalable or biased. Effective science advocacy therefore requires tools that prioritize evidence quality and transparency \cite{national2017communicating}. Retrieval-Augmented Generation (RAG) systems \cite{lewis2021retrievalaugmentedgenerationknowledgeintensivenlp}, which ground claims in sources to reduce hallucination, are integral to such solutions, yet they suffer from a critical flaw: ``methodological blindness.''

Furthermore, even existing specialized scientific claim validation models often focus on abstracts and use benchmarks like SCIFACT \cite{wadden-etal-2020-fact} or restrict analysis to abstract-level argumentation \cite{freedman2024detecting}, prioritizing dataset scale over full-text inference. Similarly, validation based on citation patterns or reputational signals \cite{chen2025pubguard} fails to audit internal methodological rigor, making these systems unreliable for advocacy decisions.  

While modern LLMs can detect obvious flaws, relying on world-knowledge reasoning diminishes transparency. We view our problem as post-retrieval methodological vulnerability detection: given a plausible study summary and its source paper, the system must decide whether the summary hides methodological weaknesses that should trigger additional human scrutiny. In research ethics, vulnerability is treated as a context-dependent increased risk of harm rather than a simple binary label \cite{gordon2020vulnerability}. Analogously, we use ``methodological vulnerabilities'' to describe summaries whose underlying evidence makes downstream audiences more likely to be misled if amplified uncritically. By employing Small Language Models (SLMs) in a structured two-agent process, the system first audits source papers against our proposed \textit{Veritable} taxonomy of scientific rigor, and subsequently synthesizes findings into a Toulmin argument structure. This separation ensures that detected weaknesses are explicitly linked to specific claims, enforcing focused, step-by-step auditing while providing the transparency required for high-stakes science advocacy. Using SLMs
 enables affordable, private, and consistent anal
ysis. 
\begin{figure}[t]
    \centering
    \includegraphics[width=\linewidth]{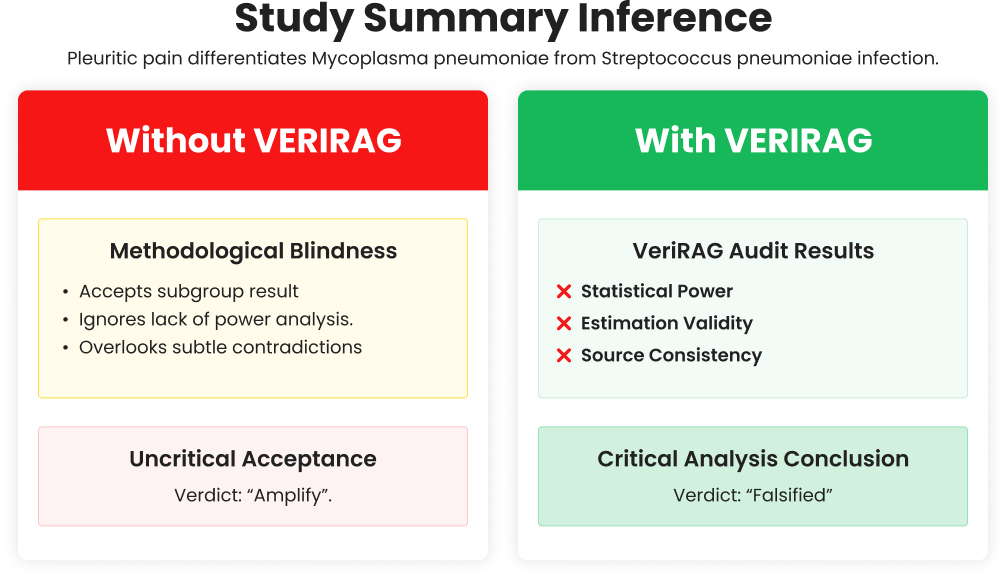}
\caption{Methodological blindness in RAG: standard systems accept flawed evidence; \textsc{VeriRAG} identifies weaknesses through structured audit.}
\label{fig:figure1}
\end{figure}
We evaluate \textsc{VeriRAG} on a seed dataset of summaries faithful to real-world retracted papers, and a primary benchmark of 1,730 summaries derived from recent literature, comprising both faithful instances and those containing realistic, non-obvious perturbations. Results across three SLM architectures demonstrate that our structured audit yields substantial performance gains without relying on the opaque reasoning of frontier models. Furthermore, human validation confirms the practical utility of our system: our test users found the generated audit trails effective for verification even when automated verdicts required nuance, validating \textsc{VeriRAG}'s role as a human-in-the-loop auditing tool.

\section{Related Work}
\label{sec:related_work}

\begin{figure*}[t]
    \centering
    \includegraphics[width=\textwidth]{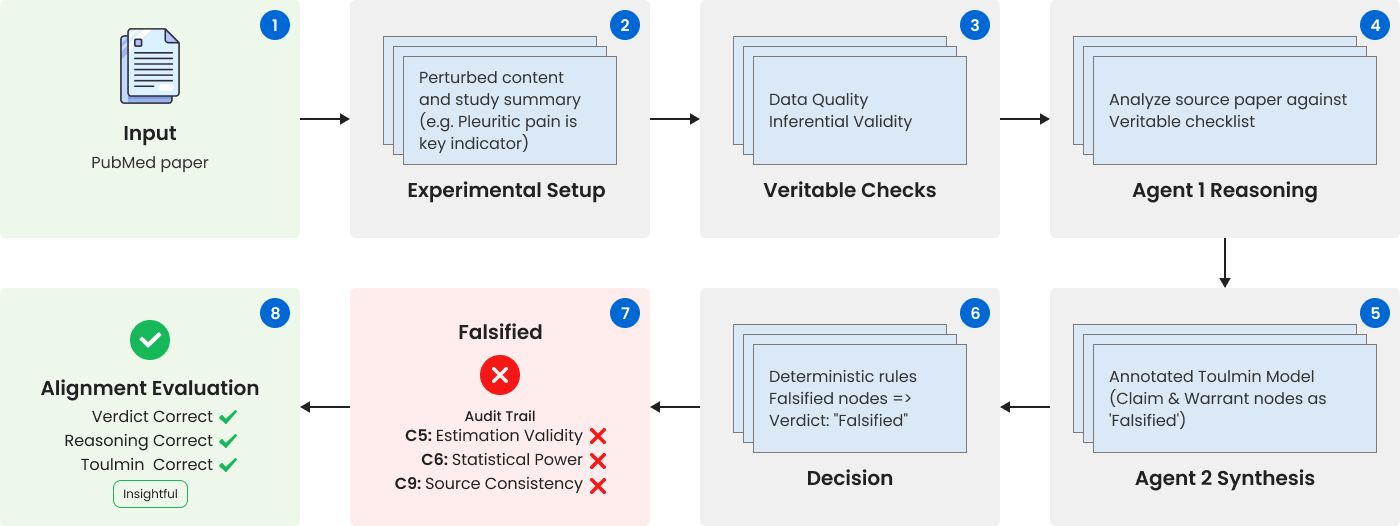}
\caption{\textsc{VeriRAG} workflow: (1) Input summaries; (2) Load Veritable; (3) Structure prompts; (4) Agent~1 audits; (5) Agent~2 synthesizes; (6) Apply decision rules; (7) Generate verdict and audit trail; (8) Evaluation}
\label{fig:verirag_workflow}
\end{figure*}

\textsc{VeriRAG} builds upon RAG and structured reasoning research, addressing core challenges in automated scientific verification.

\subsection{Reasoning Structure, Consistency, and Auditability} Approaches like Chain-of-Thought \cite{wei2023chainofthoughtpromptingelicitsreasoning} and Stepwise-FV \cite{vladika2025stepwise} decompose complex reasoning but lack formal methodological assessment structures. Reflective frameworks like Self-RAG \cite{asai2023selfraglearningretrievegenerate} and CRAG \cite{yan2024correctiveretrievalaugmentedgeneration} use self-critique; CalibRAG \cite{jang2025reliable} aligns model confidence with correctness. Trustworthy systems require transparency \cite{ifcn2016code}, yet textual rationales provide weak audit trails. \textsc{VeriRAG} addresses these through Agent~1's structured Veritable analysis, Agent~2's Toulmin synthesis, cross-SLM consistency analysis, and comprehensive audit trails.

\subsection{Evidence and Query Quality} RAG systems are vulnerable to flawed, noisy, or malicious documents \cite{shen2025reliabilityrag}. ReliabilityRAG uses reliability signals and graph methods to filter corrupted sources; AstuteRAG \cite{wang2024astuterag} consolidates sources for knowledge conflicts; VeriMinder \cite{mohole2025veriminder} detects query vulnerabilities. \textsc{VeriRAG}'s Veritable checks score evidence by methodological rigor rather than relevance, while structured prompts elicit unbiased analysis.

\subsection{Scientific Framing for Public Communication} Scientific communication shapes public perception per the Social Amplification of Risk Framework \cite{kasperson1988social}. Amplification stations (media, policymakers, networks) filter and reframe technical assessments for public understanding. \textsc{VeriRAG} supports gatekeepers with rigorous methodological assessments for informed amplification decisions.

Unlike approaches addressing individual challenges, \textsc{VeriRAG} provides a unified framework specifically for methodological assessment supporting information gatekeepers (Figure~\ref{fig:verirag_workflow}).

\section{Methodological Assessment}
\label{sec:framework}

Our framework assesses whether study summaries align with methodological rigor in source documents through structured analysis rather than semantic matching. This involves defining summary components, applying formal audit criteria, synthesizing findings using an argument model, and applying deterministic decision logic.

\subsection{Study Summary Representation}

A study summary represents a structured argument containing: Context Setting (background/research question), Evidence Nuggets (key results and data points), Atomic Entailments (decomposed factual statements), and Inference (primary conclusion). This structure enables component-wise validity analysis and facilitates tracing which specific evidence supports which claims (see Example~\ref{ex:hcq} for an illustration of this structure). This structure also overcomes the forced, claim level distillation of complex scientific studies which is rarely sufficient or relied upon by science communicators.

\begin{examplebox}{Selection Bias (Hydroxychloroquine)}
\label{ex:hcq}
\small
\textbf{Context Setting:} The study aimed to assess the effect of hydroxychloroquine treatment on the duration of viral shedding in COVID-19 patients.

\textbf{Evidence Nuggets:}
\begin{itemize}[leftmargin=*]
    \item At day 6 post-inclusion, 70.0\% (14 out of 20) of patients treated with hydroxychloroquine had a negative nasopharyngeal PCR test.
    \item In contrast, only 12.5\% (2 out of 16) of patients in the control group had a negative PCR test at day 6.
    \item The difference in virological cure between the hydroxychloroquine and control groups was reported as statistically significant, with a p-value of 0.001 at day 6.
    \item In a subgroup of 6 patients treated with both hydroxychloroquine and azithromycin, 100\% achieved virological cure by day 6.
\end{itemize}

\textbf{Atomic Entailments:}
\begin{itemize}[leftmargin=*]
    \item A group of 20 COVID-19 patients was treated with hydroxychloroquine.
    \item A control group of 16 COVID-19 patients was not treated with hydroxychloroquine.
    \item The proportion of patients with negative PCR results was measured daily for 6 days.
    \item A significantly higher percentage of patients in the hydroxychloroquine group tested negative for the virus by day 6 compared to the control group.
\end{itemize}

\textbf{Inference:} Hydroxychloroquine treatment is significantly associated with viral load reduction/disappearance in COVID-19 patients and its effect is reinforced by azithromycin.
\end{examplebox}

\subsection{Evaluation Foundations}

We draw on established methodologies: Information Nuggets from TREC QA \cite{voorhees2003overview} identify essential atomic facts for fine-grained assessment \cite{pradeep2025great}, while Entailment Trees \cite{dalvi2022explaining} map multi-premise reasoning from facts through intermediate conclusions to final hypotheses. \textsc{VeriRAG}'s two-agent process captures this: Agent~1 identifies evidence related to Veritable checks; Agent~2 connects findings to argument structure.
\subsection{The Veritable Taxonomy}

\begin{figure}[htbp]
    \centering
    \includegraphics[width=\linewidth]{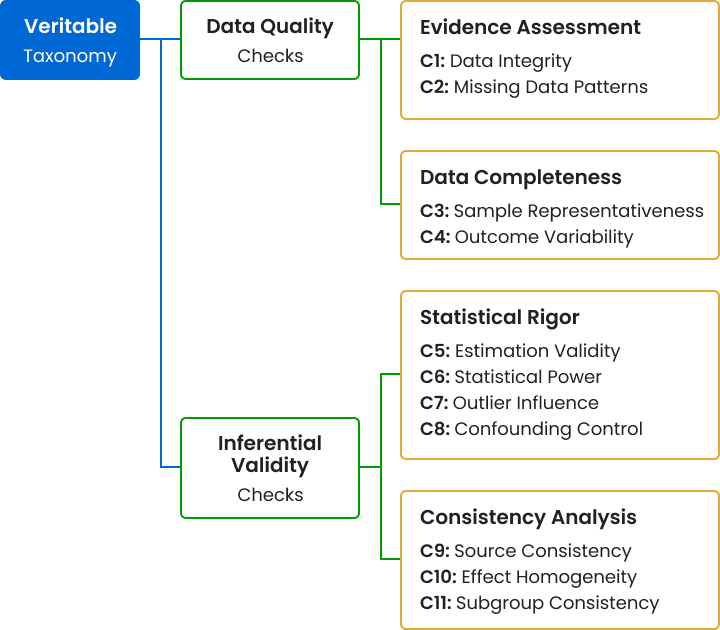}
\caption{Veritable taxonomy organized by evidence type.}
\label{fig:veritable_taxonomy}
\end{figure}

The \textit{Veritable} taxonomy is an 11-point checklist guiding methodological assessment (see Figure~\ref{fig:veritable_taxonomy}). Agent~1 systematically analyzes evidence using two categories: Data Quality (C1-C4) evaluates internal consistency, missing data patterns, sample representativeness, and outcome variability reporting; Inferential Validity (C5-C11) assesses statistical test appropriateness, sample size justification, outlier robustness, confounding control, selective reporting, heterogeneity handling in meta-analyses, and post-hoc subgroup analysis validity.

These checks map best practices from CONSORT \cite{consort2025}, STROBE \cite{vonelm2008strobe}, and PRISMA \cite{page2021prisma} guidelines to specific analysis points. For example, C1 (Data Integrity) flags inconsistencies in participant flow that might indicate data manipulation, while C9 (Source Consistency) detects when a summary emphasizes a secondary outcome while downplaying non-significant primary results. An example check definition, which maps to STROBE guidelines for observational studies, is shown below.

\begin{tcolorbox}[breakable, colback=gray!5, colframe=gray!75!black, title=\textbf{ Missing Data Patterns}]
	\begin{description}
		\item[Definition:] Analyzes how missing data is handled, with a focus on patterns that could introduce bias (i.e., non-random attrition) \cite{rubin1976inference}.
		\item[Specific Checks:] \small{\textbf{CONSORT 21c}: How missing data were handled in the analysis \textbf{CONSORT 22b}: For each group, losses and exclusions after randomisation, together with reasons}
	\end{description}
\end{tcolorbox}

\begin{tcolorbox}[breakable, colback=gray!5, colframe=gray!75!black, title=\textbf{Confounding Control}]
	\begin{description}
		\item[Definition:] For observational studies, evaluates whether key confounding variables were measured and appropriately adjusted for in the analysis \cite{pearl2009causality}.
		\item[Specific Checks:] \small{\textbf{STROBE 7}: Clearly define all outcomes, exposures, predictors, potential confounders, and effect modifiers. Give diagnostic criteria, if applicable.}
	\end{description}
\end{tcolorbox}

\subsection{Synthesis via Simplified Toulmin Model}

Agent~2 synthesizes Agent~1's findings using a simplified Toulmin model \cite{toulmin1958uses,gupta2024harnessing}, mapping summary components and Veritable findings onto core elements: Claim (main conclusion), Data/Grounds (supporting evidence), Warrant (reasoning linking data to claim), Qualifier (degree of certainty), Rebuttal (acknowledged limitations), and Backing (warrant support). Each element is assessed as "Strong (supported), "Weak (undermined by methodological concerns), or "Falsified (contradicted by critical flaws). This structured representation allows the decision agent to apply deterministic rules: if any components are marked "Falsified," the verdict is "Refutes."

\subsection{Decision Agent and Audit Trail}
The deterministic decision agent applies a strict falsification rule to the synthesized Toulmin structure: if \textit{any} node is assessed as ``Falsified'' by Agent~2, the system renders a verdict of ``Refutes.'' This binary logic prioritizes sensitivity to fatal methodological flaws over partial semantic agreement. The final output combines this verdict with the annotated Toulmin graph, creating a transparent audit trail that visualizes exactly where the scientific argument collapsed.
\section{Benchmark Dataset}
\label{sec:dataset_experiments}
Our experimental design addresses four interconnected research questions: \textbf{Q1 (Realistic Dataset):} Can we construct a dataset modeling real-world ``non-obvious'' methodological vulnerabilities where summaries are semantically fluent yet unsupported? \textbf{Q2 (System Performance):} How does \textsc{VeriRAG} perform compared to established RAG approaches in detecting these hidden vulnerabilities and mitigating methodological blindness? \textbf{Q3 (Cross-Model Robustness):} Do \textsc{VeriRAG}'s structured components generalize across different SLMs? Q4 (Human Utility): Is the generated audit trail useful to human gatekeepers as a diagnostic copilot, independent of the final automated verdict?
\subsection{Perturbation Taxonomy Development}
After reviewing the Retraction Watch \cite{retractionwatch2025} history and analyzing patterns in  retracted papers, we categorized possible perturbations into four primary types designed to model realistic scientific weaknesses rather than obvious hallucinations. Each perturbation type maps to common issues found in retracted literature:
\subsection{Dataset Construction Pipeline}
\subsubsection{Source Paper Selection and Processing} We downloaded papers published on PubMed during October 18--20, 2025, with page lengths of 15 or fewer to accommodate SLM context lengths. This timeframe ensures the papers were unseen during SLM training, preventing memorization effects. The initial collection yielded 229 papers. These papers represent a diverse cross-section of biomedical research, spanning interventional studies ($\approx$ 37\%), observational studies ($\approx$ 39\%), reviews and meta-analyses ($\approx$ 4\%), and other research types ($\approx$ 18\%), ensuring our benchmark generalizes across multiple study designs and methodological approaches. 
Papers were processed using Google DocumentAI \cite{googledocumentai} for semantic chunking, with each chunk validated by an LLM to ensure coherence and completeness. Three papers failed the chunking validation due to complex multi-column layouts and were excluded, yielding a final corpus of 226 papers.

\begin{tcolorbox}[breakable, colback=gray!5, colframe=gray!75!black, title=\textbf{Categories of Study Summaries}]
\small
\begin{itemize}
    \item \textbf{Goalpost Shifting:} Reframes secondary outcomes as primary findings while omitting non-significant primary results. Tests whether systems verify claimed findings align with study objectives.
    \item \textbf{Number Perturbation:} Subtly modifies data tables (narrowing CIs, inflating p-values) and rewrites results sections to match. Perturbed chunks replace originals during evaluation. Tests numerical consistency validation.
    \item \textbf{NLI Flips:} Distorts logical relationships to create plausible but unsupported claims, specifically via (a)~\textit{Causality Upgrade}: transforming correlational findings into causal assertions; or (b)~\textit{Overgeneralization}: extending specific sample results to broader populations.
    \item \textbf{Valid-Misleading:} Factually valid findings framed in an exclusively negative or limitation-heavy context (e.g., ``False Negative'' traps). We generate only \textit{one} per paper.
    \item \textbf{Faithful (Valid Class):} Accurately represents source papers without unsupported claims; serves as the positive class.
\end{itemize}
\end{tcolorbox}
\subsubsection{Dual-Stream Summary Generation} For each of the 226 benchmark papers, we use Gemini 2.5 Pro LLM model to generate a total of 8 summaries to ensure balanced coverage of validity types. This set includes: (1)~One Faithful Summary (Valid) that accurately reflects the paper's findings; and (2)~Seven Perturbed Summaries (Invalid). The perturbed set comprises two distinct instances each of \textit{Goalpost Shifting}, \textit{Number Perturbation}, and \textit{NLI Flips} (using differing modes, e.g., ``Causality Upgrade'' vs. ``Overgeneralization,'' to ensure diversity), plus a single instance of the \textit{Valid-but-Misleading} perturbation. This systematic generation produced an initial corpus of 1,808 candidate summaries ($226 \times 8$) before validation filtering.
\subsection{Three-Stage Quality Assurance}
\begin{figure}[H]
    \centering
    \includegraphics[width=0.75\linewidth]{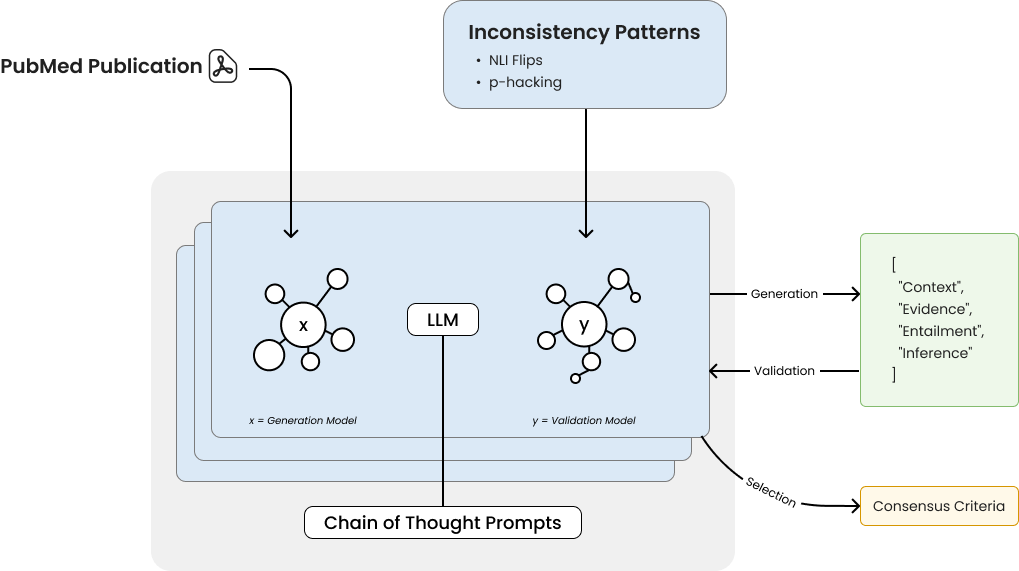}
\caption{Summary generation workflow: dual LLM generation and validation}
\label{fig:evidence_extraction}

\end{figure}
To ensure our dataset represents genuinely challenging cases requiring retrieval-based reasoning rather than surface-level detection, we applied rigorous three-stage filtering:
\subsubsection{Stage 1 Automated Open-Book Audit} We employed Claude 4.5 Sonnet \cite{claude4.5sonnet} as an independent auditor, providing it with both the exact prompt given to the generative agent and the response received. The auditor verified that: (1)~Faithful summaries accurately represented the paper; (2)~Perturbed summaries successfully implemented the intended perturbation without being trivially detectable. Out of 1,808 candidate summaries, 1,730 summaries achieved \texttt{AUDIT\_PASSED} status and were retained (Figure~\ref{fig:evidence_extraction} shows this filtering process). The 78 failures primarily involved number perturbations (34 cases) where modifications were too obvious or contradicted multiple sections, plus summaries where the perturbation was insufficiently realistic or the faithful summary contained inaccuracies. This stage ensures that our ``Invalid'' class contains genuinely subtle flaws requiring methodological reasoning to detect.

\subsubsection{Stage 2 Closed-Book Realism Validation} To confirm the dataset's ``non-obvious'' quality, we evaluated summaries using three frontier LLMs (DeepSeek-R1 \cite{deepseek2025r1}, Llama-3.1-405B \cite{meta2024llama}, Qwen3-235B \cite{qwen2024qwen3}) \textit{without} providing access to source papers. These models were prompted to assess summary validity based solely on the summary text and general scientific knowledge. As shown in Table~\ref{tab:closed_book_perf}, all models performed at level (Macro F1: 0.225--0.305), confirming that summaries appear scientifically plausible on surface inspection and require retrieval-grounded analysis to identify flaws. This realism validation provides empirical evidence for Q1: our dataset successfully models ``non-obvious'' weaknesses that create a challenging benchmark.

\begin{table}[h]
\centering
\caption{Closed-Book Frontier LLM Performance}
\label{tab:closed_book_perf}

\footnotesize
\begin{tabular}{@{}lcccc@{}}
\toprule
Model & Match & Mismatch  & Macro F1 \\
\midrule
Qwen3-235B & 468 & 1262 & 0.270 \\
Llama-3.1-405B & 532 & 1198 & 0.305 \\
DeepSeek-R1 & 391 & 1339  & 0.225 \\
\bottomrule
\end{tabular}
\end{table}

\subsubsection{Stage 3 Manual Validation on Subset} 
\begin{figure}[htbp]
    \centering
    \includegraphics[width=\columnwidth]{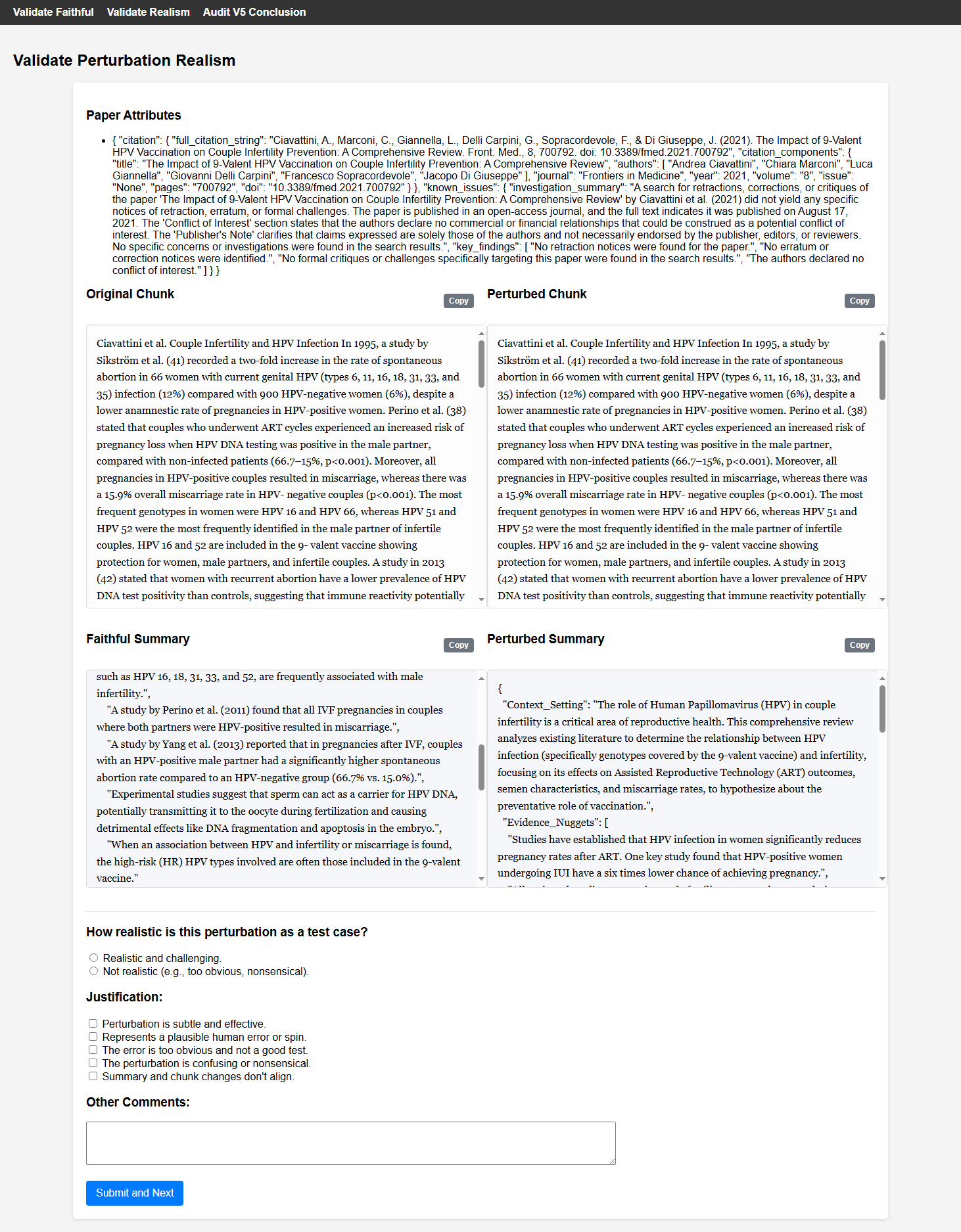}
    \caption{Interface used for manual validation of perturbation realism.}
    \label{fig:ui_realism}
\end{figure}

We built a user interface for human evaluators to review \texttt{AUDIT\_PASSED} summaries. Reviewers validated that LLM-generated study summaries maintained fidelity to source papers (for faithful summaries) and that perturbations were realistic yet detectable with careful analysis (for perturbed summaries). We conducted detailed manual checks on five papers covering different perturbation types and study designs. These summaries passed our quality checks, confirming that the automated filtering successfully identified high-quality candidates. The evaluation interface is shown in Figure~\ref{fig:ui_realism}).

\subsubsection{Stage 4 Seed Validation on Retracted Papers} To validate that our synthetic perturbations mirror real-world misconduct, we constructed a seed dataset of 30 papers from Retraction Watch \cite{retractionwatch2025}. After manually redacting retraction notices, LLM categorization revealed flaws aligning with our taxonomy: 17 data irregularities (Number Perturbations), 9 logical inconsistencies (NLI Flips), 2 misleading emphasis (Goalpost Shifting), and 2 valid-but-misleading framing (Valid-Misleading category). This validates that our perturbation taxonomy captures real patterns of scientific weakness in retracted literature.

\subsection{Baseline Implementations}

To address Q2, we benchmark \textsc{VeriRAG} against four established RAG frameworks. To ensure a fair comparison that isolates reasoning logic from model capability, we implement all baselines using identical SLM backends (GPT \cite{openai2025gptoss120bgptoss20bmodel}, Mistral \cite{mistral2025small}, and Gemma \cite{google2024gemma}) and have them receive identical evidence. We employ structured prompt chains rather than custom training, consistent with real-world deployment trends in the evolving landscape of language models.

To address \textbf{Q2}, we compare \textsc{VeriRAG} against four established RAG frameworks. The \textbf{Role-Playing Prompt} \cite{kong2024better} uses single-turn prompting where the SLM acts as a ``Scientific Analyst,'' cross-referencing claims against evidence to output a verdict with step-by-step reasoning. \textbf{Self-RAG} \cite{asai2023selfraglearningretrievegenerate} implements a two-turn process where Turn~1 generates structured critiques of evidence relevance and support, which Turn~2 synthesizes into a final verdict. \textbf{FLARE} \cite{jiang2023activeretrievalaugmentedgeneration} employs a conditional two-turn process, where Turn~1 assesses if additional information is needed and Turn~2 provides expanded evidence if requested before rendering a decision. Finally, \textbf{CER (Claim-Evidence-Reasoning)} \cite{barone2025cer} uses a two-turn baseline where Turn~1 synthesizes evidence into a prose description, and Turn~2 extracts sentiment-bearing phrases (e.g., ``Strongly Supports'') mapped to a final verdict.

We exclude a human baseline since our dataset models scenarios where methodological flaws successfully bypassed peer review, establishing a potential detection floor of 0\%. Additionally, while resource constraints precluded running frontier models on the full benchmark, we conducted a pilot on our 30-paper  seed (retracted paper) dataset using unstructured reasoning prompts. Despite generating extensive chain-of-thought rationales, performance remained low and comparable to SLM baselines: \texttt{Llama-3.1-405B} detected only 2 flaws, \texttt{Qwen3-235B} detected 4, and \texttt{DeepSeek-R1} detected 8. This suggests that unstructured reasoning in frontier models does not inherently solve the methodological blindness addressed by our framework.

\section{System Testing and Results}
\label{sec:results}

We evaluate \textsc{VeriRAG} on two distinct datasets: (1)~A real-world case study on retracted papers to demonstrate practical applicability; (2)~Our benchmark dataset of 1,730 \texttt{AUDIT\_PASSED} summaries for comprehensive quantitative evaluation. All subsequent ablations and detailed analyses are conducted exclusively on the benchmark dataset.

\subsection{Detecting Flawed Published Research}

Evaluated on our seed dataset of 30 retracted papers, \textsc{VeriRAG} correctly identified methodological flaws in 17 cases (56.7\%). In contrast, baselines overwhelmingly accepted valid-sounding but retracted science: the Role-Playing Prompt identified only 3 (10.0\%), while others performed marginally better but still missed the majority (CER: 10, 33.3\%; Self-RAG: 9, 30.0\%; FLARE: 8, 26.7\%; CHECKLIST-RAG: 5, 16.7\%). This confirms \textsc{VeriRAG}'s utility in detecting real-world scientific misconduct.  We curated this seed set by filtering Retraction Watch for papers where the primary cause for retraction was a methodological or statistical error detectable from the text. Unlike our benchmark papers which yield eight variations, we generated only one summary per redacted paper. These summaries are \textit{semantically faithful} to the source text (accurately reporting the authors' stated conclusions) but are assigned an Invalid ground truth label. This unique ``Faithful-but-Invalid'' configuration tests a system's ability to reject a summary that is textually supported but methodologically fatal. While this real-world sample validates ecological realism, we prioritized our larger benchmark for primary evaluation to leverage its Certain Ground Truth and Controlled Variation across perturbation types.

\subsection{Benchmark Dataset Evaluation}

We evaluate all models on the 1,730 \texttt{AUDIT\_PASSED} summaries from our benchmark. Given the dataset's deliberate class imbalance favoring detection of invalid science, we report Macro F1-Score as our primary metric. All subsequent analyses in this section utilize this benchmark dataset exclusively.

\subsubsection{Overall Performance (Q2)}

Table~\ref{tab:main_perf_macro} presents the primary results. \textsc{VeriRAG} consistently outperforms all baselines across all three SLM architectures. With GPT-based SLMs, \textsc{VeriRAG} achieves a Macro F1 of 0.496, surpassing the strongest baseline (CER) by 0.19 absolute points. This performance advantage is robust across backends, yielding gains of 0.26 points with Mistral and 0.24 points with Gemma over their respective best baselines.

\begin{table}[htbp]
\centering
\caption{Vulnerability Detection Performance (Macro F1).\textsc{VeriRAG} outperforms baselines across all SLMs.}
\label{tab:main_perf_macro}
\small
\begin{tabular}{@{}lccc@{}}
\toprule
Model & GPT & Mistral & Gemma \\
\midrule
\textbf{VeriRAG} & \textbf{0.496} & \textbf{0.553} & \textbf{0.458} \\
Role-Playing Prompt & 0.259 & 0.291 & 0.128 \\
Self-RAG & 0.229 & 0.239 & 0.196 \\
FLARE & 0.237 & 0.253 & 0.160 \\
CER & 0.303 & 0.245 & 0.219 \\
\bottomrule
\end{tabular}
\end{table}

Although absolute scores remain modest this reflects progress on methodological vulnerability detection, an improvement over strong RAG-style baselines which act as ``uncritical acceptors.'' We treat methodological auditing as vulnerability detection rather than ordinary text classification. Summaries in the Invalid class contain carefully constructed, non-obvious flaws modeled after retracted papers, making the task closer to detecting hidden defects than surface-level fact errors. In this context, \textsc{VeriRAG} demonstrates meaningful success in its role as a triage copilot to surface high-risk summaries and provide structured audit trails, even though the autonomous auditing problem remains far from solved.

\subsubsection{Illustrative Case Study}

\begin{tcolorbox}[breakable, colback=gray!5, colframe=gray!75!black, title=Case Study: Pneumonia Etiology Distinction]
\small
Summary Claim: ``Pleuritic pain serves as a key clinical indicator distinguishing \textit{Mycoplasma pneumoniae} from \textit{Streptococcus pneumoniae} infection.'' [\textit{Ground Truth: Invalid}]

Source: Paper reports significant difference in pleuritic pain (71.9\% vs. 41.3\%, $p=0.002$).

Baselines: Found numerical support; returned ``Supports.''

\textsc{VeriRAG}: Agent~1 identified: C5--pleuritic pain was secondary, not primary outcome; C6--small sample ($n=32$); C9--study's conclusion stated symptoms were \textit{insufficient} to distinguish pathogens (direct contradiction). Agent~2 marked \textit{Claim} as Falsified (C9 contradiction) and \textit{Warrant} as Weak (C5, C6). Verdict: Refutes.

Key Insight: \textsc{VeriRAG} distinguishes ``data exists'' from ``claim is valid''--the numerical finding was accurate, but the inference exceeded what the methodology and authors' interpretation supported.
\end{tcolorbox}

Methodological Strictness and Audit Utility: Discrepancies between \textsc{VeriRAG}'s verdicts and ground truth frequently stem from the system detecting technical inconsistencies within the source text that may not invalidate the primary claim. For instance, in a cognitive load study, Agent~1 flagged a statistical violation where repeated measures were treated as independent observations, triggering a Refutes'' verdict. Similarly, in a thyroid ablation study, the system identified data integrity issues where efficacy rates in tables did not strictly match the conclusion text. Furthermore, the precise handling of author-acknowledged limitations (the Rebuttal'' node in the Toulmin structure) remained a source of nuance. These examples reinforce our view that scientific auditing requires human intuition, positioning automated systems as assistive ``copilots'' to surface vulnerabilities rather than autonomous arbiters.

\subsection{Ablation Studies and Additional Analyses}

All subsequent ablation studies and detailed analyses are conducted on the benchmark dataset to ensure consistent evaluation conditions.

\subsubsection{Veritable vs. Model-Generated Checks}

The first ablation (CHECKLIST-RAG) tests whether expert-designed criteria outperform emergent model reasoning. Instead of the fixed Veritable taxonomy, this variant prompts the SLM to generate its own evaluation questions, then answer them to reach a verdict. Performance degradation is substantial: Macro F1 drops to 0.268 (GPT), 0.328 (Mistral), and 0.226 (Gemma), an 85\% improvement for the fixed checklist. Models frequently propose surface-level questions (``Does the study have sufficient sample size?'') but rarely generate targeted checks for confounding control, effect homogeneity, or subgroup consistency that Veritable systematically addresses. This reinforces that domain expertise in structured frameworks significantly enhances reasoning quality beyond general-purpose model capabilities.
\subsubsection{Two-Agent vs. Single-Agent}

The second ablation (Veritable-Only) collapses the two-agent architecture into single-turn processing. Our results show substantial degradation (Macro F1: 0.337 vs. 0.496, 47\% improvement for two-agent). Single-agent systems often identified methodological concerns but failed to connect them to specific claims, yielding incorrect verdicts. The two-agent structure enforces focused processing: Agent~1 audits evidence against Veritable checks; Agent~2 synthesizes findings into argument structure, explicitly linking weaknesses to claims. This separation reduces cognitive load and improves reasoning quality.

\subsubsection{Decision Logic Criteria}

The third ablation tests alternative decision rules for converting Toulmin assessments to verdicts. Table~\ref{tab:criteria_abl} compares our ``Any Falsified'' logic against Majority Weak/Falsified ($>50\%$ nodes) and $>1$ Falsified (two+ nodes). The ``Any Falsified'' rule's superiority supports the principle that a single critical flaw (e.g., underpowered sample, uncontrolled confounding) suffices to question validity.

\begin{table}[h]
\centering
\caption{Decision Logic Ablation (Macro F1)}
\label{tab:criteria_abl}
\small
\begin{tabular}{@{}lccc@{}}
\toprule
Decision Criteria & GPT & Mistral & Gemma \\
\midrule
Any Falsified & 0.496 & 0.553 & 0.458 \\
Majority Weak/Falsified & 0.353 & 0.479 & 0.394 \\
$>1$ Falsified & 0.278 & 0.375 & 0.243 \\
\bottomrule
\end{tabular}
\end{table}

\subsection{Components and Consistency}

To address Q3, we analyzed whether \textsc{VeriRAG}'s structured components guide different SLMs toward consistent reasoning on the benchmark dataset. We evaluated GPT and Mistral on the full 1,730-summary dataset, comparing: (1)~Final verdict agreement; (2)~Veritable checks identified by Agent~1; (3)~Toulmin role assignments by Agent~2; (4)~Node assessment distributions.

\subsubsection{Reasoning Distribution} Table~\ref{tab:assessment_distribution} highlights significant divergence in evidentiary standards across models. GPT exhibits a distinct leniency bias (64.3\% Strong nodes), whereas Mistral adopts a far more critical stance (highest Weak/Falsified rates at 54.0\% combined). This granular discrepancy explains the variance in final verdicts: while models verify the same structure, they apply markedly different thresholds of rigor when validating individual claims.

\begin{table}[h]
\centering
\caption{Distribution of Node-Level Assessments}
\label{tab:assessment_distribution}
\footnotesize
\begin{tabular}{@{}p{1cm}ccc@{}}
\toprule
Assessment & GPT & Mistral & Gemma \\
\midrule
Strong & 6635 (64.3\%) & 6184 (46.1\%) & 5902 (53.7\%) \\
Weak & 2483 (24.1\%) & 5026 (37.5\%) & 4060 (36.9\%) \\
Falsified & 1201 (11.6\%) & 2210 (16.5\%) & 1030 (9.4\%) \\
\bottomrule
\end{tabular}
\end{table}

\subsubsection{Component-Level Consistency:} Despite moderate verdict agreement, \textsc{VeriRAG}'s internal reasoning components show substantially higher cross-model consistency (Table~\ref{tab:component_consistency}):

\begin{table}[h]
\centering
\caption{Cross-Model Consistency}
\label{tab:component_consistency}
\resizebox{\columnwidth}{!}{%
\begin{tabular}{@{}lcc@{}}
\toprule
\textbf{Component} & \textbf{Jaccard} & \textbf{Overlap} \\
\midrule
Agent 1: Veritable Checks & 0.3383 & 99.7\% \\
Agent 2: Toulmin Roles & 0.5056 & 100.0\% \\
\bottomrule
\end{tabular}%
}
\end{table}

The high Jaccard similarity for Toulmin role assignments (0.5056 with perfect non-zero overlap) demonstrates that Agent~2's structural synthesis converges across models even when Agent~1 identifies different specific Veritable checks. This suggests the Toulmin framework successfully guides diverse SLMs toward shared structural understanding, supporting system reliability beyond individual model implementations.

These findings support Q3: \textsc{VeriRAG}'s structured framework provides sufficient constraint to guide diverse SLMs toward consistent structural analysis while allowing flexibility for model-specific interpretations of evidence strength.

\subsection{Human Validation of Audit Trail (Q4)}

To address Q4, we evaluated whether \textsc{VeriRAG}'s audit trails support practical decision-making by human gatekeepers. Two independent auditors (researchers familiar with scientific methodology) reviewed 45 balanced summaries from the benchmark (5 Correct + 5 Incorrect for each perturbation type, plus 5 Valid→Invalid cases), assessing whether the structured breakdown (Veritable findings, Toulmin graph) was ``Useful and Auditable'' for determining paper validity, regardless of the automated verdict. The evaluation interface presented the study summary, full paper, \textsc{VeriRAG} analysis, and final verdict.

\begin{figure}[h]
    \centering
    \includegraphics[width=0.8\columnwidth]{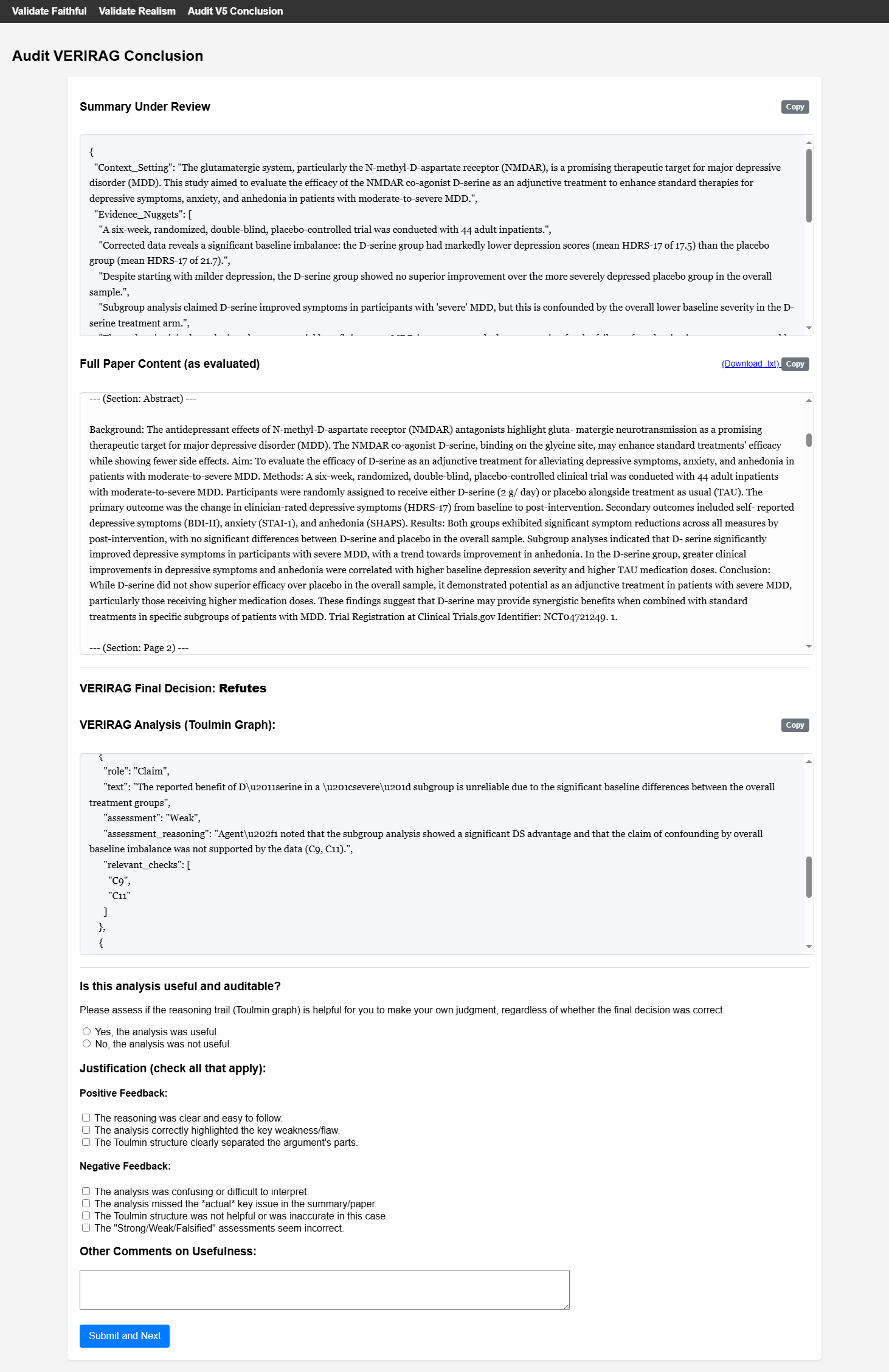}
    \caption{Interface used to evaluate  audit trail.}
    \label{fig:ui_audit}
\end{figure}
Table~\ref{tab:audit_agreement_combined} shows high inter-rater agreement (84.4\%) on overall utility. Critical Finding: When \textsc{VeriRAG}'s automated verdict was incorrect, auditors still found the reasoning trail useful in 79.2\% of cases (19/24), validating that structured audit trails support human oversight even when automated decisions fail. This directly addresses concerns about automated fact-checking: \textsc{VeriRAG}'s transparent approach ensures that even failures provide actionable information rather than opaque black-box decisions, confirming successful support for the human-in-the-loop workflow essential for responsible science advocacy.

\begin{table}[h]
\centering
\caption{Audit Trail Usefulness}
\label{tab:audit_agreement_combined}
\resizebox{\columnwidth}{!}{%
\begin{tabular}{@{}lcccc@{}}
\toprule
\textbf{Condition} & \textbf{Both Useful} & \textbf{Disagree} & \textbf{Both Not} & \textbf{Agreement ($\kappa$)} \\
\midrule
Overall (N=45) & 34 & 7 & 4 & 84.4\% (0.44) \\
Incorrect Verdicts (N=24) & 16 & 5 & 3 & 79.2\% (0.41) \\
\bottomrule
\end{tabular}%
}
\end{table}

\section{Conclusion and Future Work}
\label{sec:conclusion}

We presented \textsc{VeriRAG}, a framework establishing post-retrieval Auditing as a critical component of high-stakes RAG systems. By proving that semantic relevance is insufficient for scientific truth, we demonstrate that future RAG architectures must include an explicit auditing layer to prevent the amplification of methodologically flawed but semantically fluent studies.  On a benchmark of 1,730 summaries modeling realistic scientific flaws, our approach yields robust improvements of 0.19--0.26 absolute F1 points over standard RAG baselines across multiple architectures. These results, coupled with high human agreement on audit trail utility, confirm that structured verification provides the necessary interpretability for responsible science advocacy.

Future work will extend \textsc{VeriRAG} by incorporating causal inference techniques \cite{allein2025assessing} and confidence calibration \cite{jang2025reliable} to refine the detection of subtle flaws. Furthermore, given that many retractions stem from image manipulation, we plan to expand the audit scope to include multi-modal analysis of figures. Finally, developing interactive interfaces for granular user queries remains a key priority.

\section{Limitations}
\label{sec:limitations}

While \textsc{VeriRAG} advances automated study summary verification, several limitations frame future work avenues.

\subsection{Multi-Modal and Coverage Limitations} The framework is text-based and does not analyze figures or charts which is critical given evidence is often presented visually. Extending audit mechanisms to multi-modal content and correspondingly expanding checklist coverage remain priorities.

\subsection{Latency and Computational Cost} The two-agent architecture and full paper content audit inherent to \textsc{VeriRAG} introduces a latency trade-off to achieve methodological depth. Our system requires a median of 25.7 seconds per summary using GPT-based SLMs. This represents a latency increase of approximately 20 seconds compared to simple single-turn role-playing prompts ($\approx$ 5.9s) and reflects the additional computational overhead of the structured two-agent reasoning compared to a single-turn Veritable checklist ($\approx$ 10.4s). This duration stems from the sequential nature of the detailed audit and the detailed output generation required for the Toulmin synthesis, further compounded by network overhead from the use of \texttt{Together.ai} serverless endpoints. However, we consider this response time acceptable for asynchronous ``human-in-the-loop'' auditing workflows, where accuracy and the generation of a transparent audit trail is important.

\subsection{SLM Reasoning and Mapping Challenges} Despite structured guidance, SLMs remain susceptible to hallucination, misinterpretation of complex statistical concepts, or failing to detect subtle flaws \cite{ji2023survey}. Agent~2's mapping of nuanced findings onto discrete Toulmin components (Strong/Weak/Falsified) can oversimplify complex situations or struggle with conflicting evidence. While the two-agent structure mitigates risks compared to free-form reasoning, error propagation remains possible.

\subsection{Generalization and Prompt Sensitivity}
Although \textsc{VeriRAG} employs structured intermediate representations to stabilize outputs, prompt sensitivity remains an inherent challenge in LLM-based reasoning \cite{gupta2024harnessing,allein2025assessing}. Furthermore, while our evaluation establishes efficacy within the critical and complex domain of biomedical literature, extending this validation to disparate scientific fields or non-academic genres requires further empirical investigation.

\begin{acks}
This work was performed under the auspices of the U.S. Department
of Energy by Lawrence Livermore National Laboratory under Contract
DE-AC52-07NA27344 and was supported by the LLNL-LDRD Program under Project No. 25-SI-001. LLNL-JRNL-2013879-DRAFT.
\end{acks}
\bibliography{tacl2021}
\bibliographystyle{acl_natbib}

\end{document}